# MultiGBS: A multi-layer graph approach to biomedical summarization

Ensieh Davoodijam
*Department of Electrical and Computer Engineering,
Isfahan University of Technology,
Isfahan 84156-83111, Iran*
`e.davoodijam@ec.iut.ac.ir`

Nasser Ghadiri *
*Department of Electrical and Computer Engineering,
Isfahan University of Technology,
Isfahan 84156-83111, Iran*
`nghadiri@iut.ac.ir`

Maryam Lotfi Shahreza
*Department of Computer Engineering,
University of Shahreza,
Shahreza 86149-56841, Iran*
`lotfi@shahreza.ac.ir`

Fabio Rinaldi
*Swiss AI Lab IDSIA / Istituto Dalle Molle di Studi sull'Intelligenza Artificiale
CH-6962 Lugano, Switzerland*
`fabio.rinaldi@idsia.ch`

## Abstract

Automatic text summarization methods generate a shorter version of the input text to assist the reader in gaining a quick yet informative gist. Existing text summarization methods generally focus on a single aspect of text when selecting sentences, causing the potential loss of essential information. In this study, we propose a domain-specific method that models a document as a multi-layer graph to enable multiple features of the text to be processed at the same time. The features we used in this paper are word similarity, semantic similarity, and co-reference similarity, which are modelled as three different layers. The unsupervised method selects sentences from the multi-layer graph based on the MultiRank algorithm and the number of concepts. The proposed *MultiGBS* algorithm employs UMLS and extracts the concepts and relationships using different tools such as SemRep, MetaMap, and OGER. Extensive evaluation by ROUGE and BERTScore shows increased F-measure values.

1. **Introduction**

The amount of data being generated and collected in the biomedical sciences is rapidly increasing. The number of biomedical articles published in PubMed is currently more than 30 million, and it increases at the rate of more than 1 million per year [1]. To efficiently process such a high volume of data, it is essential to design effective mechanisms that enable the rapid extraction of information from data. Text summarization aims to identify the essential, meaningful information in a single document or a set of related documents [2]. Such summaries can then be used in other areas of research such as information retrieval, question answering(IR), and text classification [3].

In general, existing summarization approaches can be separated into five categories: (1) statistical-based, (2) topic-based, (3) discourse-based, (4) machine learning-based, and (5) graph-based methods. Although these approaches have been shown to be efficient, they suffer from performance degradation and require extensive training [4]. We discuss these challenges in greater detail in Section 2.

In recent years, graph-based approaches have received increased attention in data mining, in particular in the text summarization domain. There are two fundamental issues associated with graph-based methods:
- How should a graph be assembled?
- How should an algorithm for sentence ranking be developed?

In a graph-based approach, each node represents an object, and each edge represents the relationship between a pair of objects [5]. Basic graph algorithms for text summarization are often limited to covering just a single type of relationship, such as word similarity or semantic similarity. This limitation prevents them from capturing enriched information in the input text and covering multiple aspects or subtopics at the same time. Moreover, biomedical documents often contain some abbreviations, acronyms, and symbols that existing algorithms for graph-based approaches are usually unable to detect properly, which consequently lead to the loss of valuable information during the process of ranking the documents' sentences [5,6]. Therefore, new algorithms are needed to increase the diversity and coverage of summarized results.

In this paper, we propose MultiGBS, a novel multi-layer graph-based biomedical text summarizer that models three different types of relationships between the sentences of a given document. This system benefits from domain knowledge for extracting biomedical entities and relationships. It first builds an undirected weighted multi-layer graph from the source document. The multi-layer graph simultaneously covers semantic relationships, word relationships, and co-reference relationships. The proposed MultiGBS algorithm uses the Unified Medical Language System (UMLS) knowledge source to extract concepts and identify the correlations among them. The UMLS is a collection of several biomedical vocabularies and standards [7,8]. We use MetaMap [9] and OGER [10], which are two different tools capable of identifying concepts and map original text to UMLS concepts. Finally, the proposed method employs the MultiRank algorithm [11] on the multi-layer graph to rank the sentences of the given document. The MultiRank algorithm is similar to PageRank and runs on multi-layer graphs [11].

The system is evaluated using the Recall-Oriented Understudy for Gisting Evaluation (ROUGE) metrics [12,13] and BERTScore [14]. As the experimental results show, our algorithm achieves better results in comparison with other summarization tools.

The rest of the paper is organized as follows. Section 2 briefly reviews the history of text summarization and outlines related work. Section 3 presents a description of our biomedical text summarization method. The results are presented and discussed in Section 4. Finally, Section 5 reports conclusions and limitations.

## 2. Related Work

There are different categorization approaches for summarization methods depending on the type of input, type of output, and the user requirements. In one direction, summarization methods can be classified into *generic* versus *query-oriented* methods based on the purpose of the summaries [15]. A query-oriented process creates an outline that, unlike a general summary, includes the content related to a given query. In another categorization approach, depending on the number of source documents, the summarizers can be categorized into *single-document* versus *multi-document* methods. In the third approach, the output of the summarizer can be classified as *extractive* versus *abstractive* summaries. Extractive summarization chooses the most meaningful subset of the sentences in the document set, while abstractive summarization creates new sentences that are unseen in the sources. Another approach divides the summaries into *indicative* versus *informative* summaries. Informative summaries include sufficient content, and users do not require the original input for understanding. On the other hand, indicative summaries only provide a view of the content, and users still need to review the original document [3,15,16]. The proposed method provides a generic, single-document, informative, and extractive summary.

As mentioned in Section 1, there are five categories of extractive approaches for summary generation, which are briefly discussed below.

- **Statistical-based approaches:** These techniques consider statistical features such as the position of the sentence, positive keywords or negative keywords, the centrality of the sentence, the similarity of the sentence to the title, the length of the sentence, the presence of numerical data in the sentence, TF*IDF (Term Frequency–Inverse Document Frequency), etc. [17–19]. It has been shown that the use of statistical features alone is not sufficient, and a combination of these features with other methods could provide better results [3,16].
- **Topic-based approaches:** In these methods, summaries are created by identifying the topic. The topic is the primary concern of the document and is described in five different representation: (1) topic signatures, (2) enhanced topic signatures, (3) thematic signatures, (4) modeling the document's content structure, and (5) templates. The different representation of topics identified the relevant relation, terms and concepts to create the summary. The algorithms in this category are sophisticated and require considerable skills to use [20].
- **Machine learning-based approaches:** These methods are based on well-known machine learning algorithms such as classification [21], hidden Markov models [22], Bayesian methods, neural networks [23], support vector regression (SVR) and least angle regression [24]. SummaRuNNer is an extractive summarizer that uses a bi-directional recurrent neural network for sentence representation and sentence selection [25]. Cheng and Lapata introduced NN-SE as an extractive summarizer. In this model, sentences are represented with convolutional neural networks (CNNs), and sentences are selected with NN-SE [26]. PriorSum uses the gold standard summaries for training. It merges a multi-layer CNN with statistical features such as sentence position and average term frequency [27]. Narayan et al. [28] introduced an extractive summarization based on structured transformers. A limitation of the existing methods in this category is the need for a large training corpus [3].

- **Discourse-based approaches:** In this group, the summarization methods consider linguistic knowledge [29,30]. Afnan et al. [31] employ the rhetorical structure theory and the MEAD summarizer system to increase the coherence and correctness of summaries. Other approaches combine statistical and linguistic techniques. These algorithms have a moderate performance and do not provide significant improvements [3].

- **Graph-based approaches:** Graph-based summarization methods represent the document as a graph through nodes, which show different parts such as terms, phrases, concepts, sentences, and edges which describe the similarity relations between them [5]. There are various measures for calculating the similarity between the text units, such as cosine similarity [32], the longest common subsequences [33], and the number of common words [34].

    LexRank is one of the most well-known graph-based approaches for multi-document summarization; it first creates a weighted graph for a given document based on a predefined threshold and then finds the essential sentences of the document through a random walk on the resulting graph [32].

Some domain-independent summarization systems use frequent item-set mining [35–37]. Other methods use UMLS as a domain of knowledge resources and extract the concepts and vocabularies [6,38–41]. For instance, BioChain [42] is a typical single document summarization that uses UMLS and creates concept chains that are ranked based on concept frequency. Another method creates a simple graph from the input text based on UMLS concepts [43] and measures the similarity between nodes based on an "is-a" relationship and clusters the graph based on genetic graph clustering (GGC). Also, Nasr et al. [41] created a simple graph and used an n-gram based on frequent set mining to create edges. Most of the existing graph-based algorithms for biomedical text summarization cannot handle more than one relationship type between the text elements simultaneously, thus providing limited accuracy.

In this paper, we focus on biomedical summarization methods to extract the exact meaning from different words or terms based on the domain. The proposed method is a novel biomedical summarization method that uses multi-layer graphs instead of simple graphs, making it possible to handle several types of relationships between the sentences in the text. For this aim, MultiGBS employs different tools such as MetaMap, OGER, and SemRep to extract the concepts and relations from the biomedical domain. These tools are introduced in Section 3.

## 3. Methods

MultiGBS is a graph-based method that has two main steps: *graph creation* and *sentence selection*, as presented below.

### 3.1 Graph creation

The focus of our approach is on creating a graph based on heterogeneous network concepts, i.e., a network that involves different types of objects or links. There are two types of heterogeneous networks [44]:

- *Multimode networks* that consist of different types of objects with various relationships between them.
- *Multidimensional networks* or *multi-layer networks* that have multiple types of relationships, and each layer represents different relationships among nodes.

A high-quality summary should represent different aspects of the original text and cover as much useful information as possible by reflecting on the essential topics. Therefore, the summarization method should be able to extract different relationships from the text and concepts. Multi-layer graph is a well-known technique used to aggregate and present different types of relationships. Our proposed MultiGBS method uses multi-layer networks to cover essential aspects and different types of relationships between sentences. A document is modeled as a weighted graph in which every sentence is a node, and the different types of edges represent various relationships between sentences. A formal definition is given below:

**Definition 1.** Let *D* be a document with a set of sentences *S*. The document can be represented as an undirected weighted graph $G = (V, E, \alpha)$, where *α is* a set of layers *α=1, 2, M, and* V is a collection of vertices also called nodes, and *E* shows the quadruple *(u, v, α, w),* where *w* is the weight on edge between *u* and *v* in the layer *α*.

Our network has three layers, and each layer represents a different similarity, as shown below:
- Semantic layer
- Word layer
- Co-references layer

**Semantic Layer**: MultiGBS extracts concepts and semantic types from the UMLS. It is widely used in biomedical text summarization research as a collection of several vocabularies and standards. UMLS has three primary data sources: (1) Metathesaurus, (2) Semantic Network, and (3) Specialist Lexicon, which are briefly explained below [8].

- UMLS Metathesaurus: A set of various biomedical concepts, names, and synonyms obtained from approximately 200 different vocabularies. It is considered the major UMLS component.

- UMLS Semantic Network: All concepts in the UMLS Metathesaurus can be classified using the Semantic Network component. It defines 133 general categories and 44 relationships between categories.

- UMLS Specialist Lexicon: A database that provides lexical data and programs for language processing.

To identify and extract biomedical concepts from a text document, MultiGBS employs two different tools:

1. **MetaMap**: developed at the National Library of Medicine that maps biomedical text to the concepts of the UMLS Metathesaurus [8,9].
2. **OGER:** OntoGene's entity recognizer is a text-mining tool for identifying biomedical concepts. It combines a dictionary-based annotator with a corpus-based disambiguation component [10,45].

After extracting concepts, our summarizer measures the semantic similarity using the n-gram algorithm [46].

**Word Layer**: In this step, unlike the semantic layer, MultiGBS only considers the exact words in sentences and calculates the similarity between sentences based on n-gram.

**Co-reference Layer**: MultiGBS creates a relationship between two sentences based on co-reference resolution. It is the task of finding all expressions that refer to the same entity in a text. For example, consider this sentence:

> *My favorite sport is football because it a way of life, Sara said.*

"My" and "Sara" refer to the same entity, and "football" and "it" refer to the same entity[47]. SemRep [48] is used to extract co-reference resolution. It applies structured domain knowledge from the UMLS and obtains semantic propositions from biomedical text [48].

Finally, a document is represented as a weighted undirected multi-layer graph in which nodes are the sentences, and the edges show the different similarity relations between these sentences. The different weights show the strength of the links in the original text.

For a better illustration of the advantages of the proposed multi-layer graph architecture in capturing different relationships between sentences, an example is provided from [49]. Several parts of a MEDLINE abstract (PMID 21349396) are selected for this purpose.

Input text:

- *"Pulmonary arterial hypertension (PAH) is a rare and progressive disease of the pulmonary arterial circulation …."*

- *"There are currently <u>3 classes of drugs approved for the treatment of PAH: prostacyclinanalogues, endothelin receptor antagonists, and phosphodiesterase type 5 inhibitors. …</u>"*

- *"Although definitive evidence will require randomized and properly controlled long-term trials, the current evidence supports the long-term use of <u>these drugs</u> for the treatment of patients with PAH."*

It can be observed from the example that the abbreviation "PAH" is recognized as a disease and prostacyclin analogues, endothelin receptor antagonists and phosphodiesterase type 5 inhibitors are identified as drugs using UMLS. We were also able to consider the underlined mentions as the same concept [49]. Fig. 1 shows a multi-layer graph of the sample document.

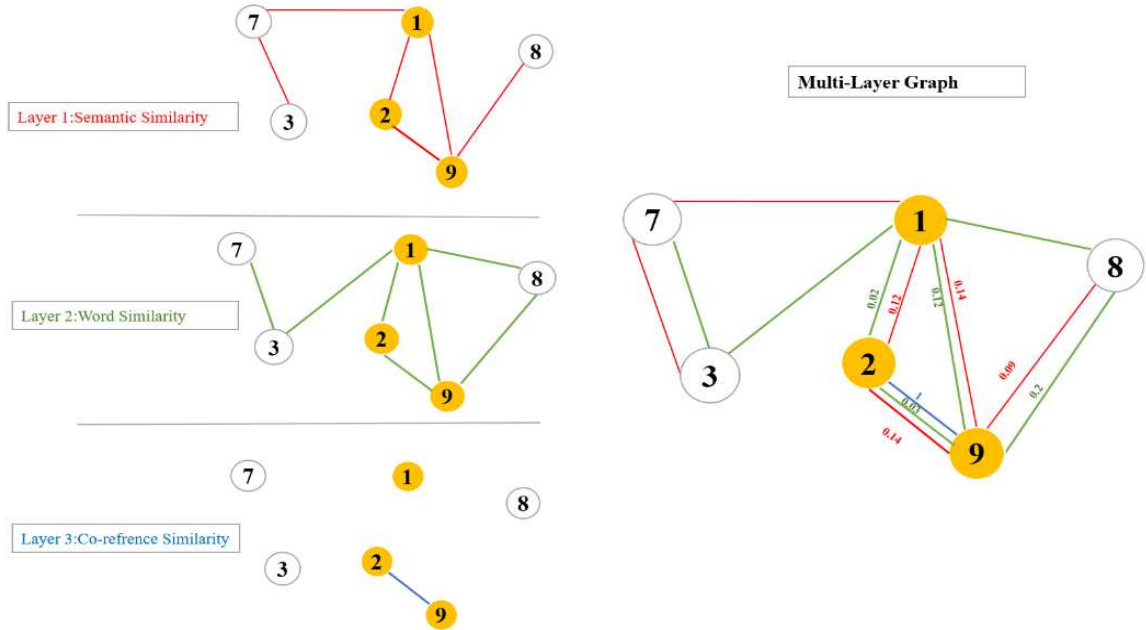

**Fig. 1**: Multi-layer graph with three layers. The red edges show the first layer (Semantic Similarity), the green ones represent the second layer (Word Similarity), and the purple edges display the third layer (Co-reference Similarity). The yellow vertices represent the three sentences of the example. (Only the parts of the sample document are represented.)

## 3.2 Sentence selection

After creating the graph, the proposed algorithm needs to select the sentences to generate the summary. As mentioned in Section 2 for the extractive summary, the summarization system requires choosing essential sentences from the original text. MultiGBS ranks all the sentences based on the multi-layer network and selects the top-ranked sentences according to the compression rate. The compression rate is the percentage of the original sentences to be shown in the summary. We use an algorithm that aggregates three similarity measures on the multi-layer graph and ranks the sentences. Therefore, we examine two different methods for sentence selection, which are explained below.

### 3.2.1 MultiGBS sentence selection algorithm – the basic version

In this step, our method requires aggregating the different similarity measures, without losing data. The MultiGBS algorithm uses MultiRank [11] to rank all the sentences. MultiRank is based on a random walk algorithm and works on multiple different layers. It discovers the centrality of nodes and the layers based on various properties. The output of this algorithm is the centrality $X_i$ of the nodes *i= 1, 2, n*, that assigns centrality values based on highly influential layers and central nodes. The MultiRank algorithm calculates three different matrices which extract from a multi-layer graph [11]. First, the MultiRank algorithm creates three networks from the multi-layer graph [11]:

- Aggregate network: the aggregate network is a single network that aggregates all layers. A link is created if there is at least one link in one layer between the nodes.
- Bipartite network: this network determines which nodes are connected in which layers.
- Colored network: In this network, the MultiRank algorithm computes the different influences associated with each layer.

After that, the centrality $X_i$ of the node i is calculated based on these networks and the relationships between them. In this version, top sentences are selected by sorting items of $X_i$ in descending order:

$$Score(S_i), \leftarrow Sorted(X_i) \tag{1}$$

### 3.2.2 MultiGBS sentence selection algorithm – the enhanced version

The MultiGBS-Enhanced algorithm considers the number of concepts for each sentence and proposes $LenCon(S_i)$ as follows:

$$LenCon(S_i) \ \text{<-} \ \text{the number of concepts } (S_i)/ \text{ the total number of concepts in the document} \tag{2}$$

We use the MultiRank algorithm to calculate $Sorted(X_i)$ (similar to MultiGBS Basic) and calculate the $Score(S_i)$ according to Eq. (3). Different weights between -1 and 1 can be assigned to $\gamma$ and $\theta$. Finally, the scores are normalized based on min-max normalization.

$$Score(S_i) \leftarrow \gamma Sorted(X_i) + \theta LenCon(S_i) \tag{3}$$

The pseudocode of MultiGBS Enhanced is shown in Algorithm 1.

**Algorithm** 1**:** MultiGBS Enhanced algorithm

```
Function Create_Graph(document) returns a Multi-Layer Graph
Input: D // a document
Local variables: S= {s₁,s₂,…,sₙ} // list of sentences
                 C= {c₁,c₂,…,cₙ} // list of concepts
                 T= {t₁,t₂,…,tₙ} // list of tokens
                 R = {r₁,r₂,…,rₙ} // list of co-references relationships
```

```
For all s_i in S do
  For all s_j in S do
      Layer1 ← Semantic_Sim (C_{S_i}, C_{S_i})
      Layer2← Word_Sim (T_{S_i}, T_{S_i})
      Layer3← CoRef_Sim (R_{S_i}, R_{S_i})
Function MultiGBS _ Sentence selection_Enhanecd (Multi-Layer Graph) returns a sorted list of
sentences
Input: I = (V, E, α)
Local variables: G, W, A // Matrix
                 W ←Create_ Aggregate_ network (I)
                 B←Create_ Bipartite_ network (I)
                 G ← Create_ Colored _ network (I)
```

$$W^{[\alpha]} = \sum_{i=1}^{N}\sum_{j=1}^{N} A_{ij}^{[\alpha]}; \; B_{\alpha i} = \frac{\sum_j A_{ji}^{[\alpha]}}{W^{[\alpha]}}; \; G_{ij} = \sum_{\alpha=1}^{M} A_{ij}^{[\alpha]} z^{[\alpha]}; \; Z^{\{\alpha\}} = \frac{1}{\omega} W^{[\alpha]} \sum_{i=1}^{N} B_{\alpha i} X_i$$

$X_i = \tilde{\alpha} \sum_{i=1}^{N} \frac{G_{ji}}{k_j} X_j + \beta v_i$ // the centrality $X_i$ of the node, i given the influences $Z^{\{\alpha\}}$ α is

taken to be $\tilde{\alpha} = 0.85$ and $k_j, v_i, \beta$ are calculated based on W, B, G

$S(X_i) \leftarrow Sort_{Asc}(X_i)$ // Sort items of $X_i$ in an descending order

```
For all v_i do
  LenCon(v_i)  <- the number of concepts (v_i)/ the total number of concept in document
  Score(v_i) ← γSorted(X_i) + θLenCon(v_i)
Normalize _min-max (Score(v_i))
Function MultiGBS -ENHANCED (D) returns a summary
Input D // a document
Local variables: Multi_Graph // a multi-layer graph
Multi_Graph= Create_Graph(D)
MultiGBS _ Sentence selection_Enhanecd (Multi_Graph)
```

## 4. Experimental evaluations

In this section, the MultiGBS algorithm is evaluated. In the following, the overall evaluation approach as well as the evaluation criteria, the baseline and the corpus are described.

### 4.1 Experiment design

There are two main categories for text summarization evaluation: intrinsic and extrinsic. Extrinsic evaluation attempts to quantify information content based on measures such as the success rate, time-to-completion, and decision-making accuracy. On the other hand, intrinsic methods evaluate the quality or informativeness [50]. A standard method for intrinsic evaluation is to compare the content provided by the summarizer with the human-produced model summaries as the reference summary. Creating a reference summary is a complex, time-consuming process that requires expert input [3,16,50]. Therefore, we used intrinsic evaluation and designed experiments incrementally. The output of different summarization methods can be found in Appendix A.

The initial experiments show how different features can affect the ROUGE scores. This step was intended to address the following research questions:

Q1) To what extent does the type of graph influence the results?

Q2) To what extent does extracting irrelevant concepts influence the results?

Q3) To what extent does the use of different entity extraction strategies influence the results?

Q4) To what extent does the compression rate influence the results?

Q5) To what extent does the subset of representation layers influence the results?

We augmented an experiment to compare the proposed MultiGBS algorithm with other methods. We also applied a statistical test to validate the results using the SciPy Python package [51]. The statistical significance was evaluated by the "Wilcoxon" signed-rank test.

### 4.2 Evaluation metrics

To measure the effectiveness of our algorithm, ROUGE and BERTScore were used. BERTScore is based on Bidirectional Encoder Representations from Transformers (BERT) [52] as a pre-trained language representation model. This score does not compare the exact words; instead, it calculates the token similarity based on BERT [14].

ROUGE is a set of metrics that compares an automatically produced summary with one or more human-made summaries. ROUGE is defined in several versions with different applications. It estimates precision, recall, and F-measure. In this work, ROUGE-1, ROUGE-2, ROUGE-L, and ROUGE-SU4 measures are used to evaluate the summarizer [12]. There is a wide range of research works on summarization that evaluated their outcomes using ROUGE based on the Document

Understanding Conferences[1] [53] and the Text Analysis Conferences[2] [54]. The BioASQ challenge for the biomedical domain also evaluates the task of automatically summarizing biomedical texts as part of a question answering [55,56]. The values of recall, precision, and F-measure are reported. As a basic example of how ROUGE works, one can consider the following example:

- System Summary 1: The book was found under the bed.
- System Summary 2: The little red book was found under the big funny bed.
- Reference Summary: The book was under the bed.

For any sequences X and Y, the longest common subsequence (LCS) of X and Y is a common subsequence with a maximum length. As described in [12], recall and precision for ROUGE-L are calculated as follows (X is a reference summary of length m, and Y is a system summary of length n):

$$R_{lcs} = \frac{LCS(X,Y)}{m} \tag{4}$$

$$P_{lcs} = \frac{LCS(X,Y)}{n} \tag{5}$$

Therefore, $R1 = R2 = 1$ and $P1 = \frac{6}{7} = 0.86$, $P2 = \frac{6}{11} = 0.55$ are calculated.

While current research works often report the recall value, the precision is also useful in expressing the irrelevant words in the generated summary [6,16,41,43,57]. Therefore, we reported the values of recall, precision, and F-measure.

## 4.3 Evaluation corpus

To the best of our knowledge, there is no widely accepted standard corpus for biomedical documents and their model summaries. Existing works commonly use a collection of biomedical articles and consider the abstract of each article as the model summary [6,16,40,43,58,59]. Therefore, we randomly selected 450 biomedical scientific articles from BioMed Central. As explained in [60], the size of the dataset is large enough to evaluate results and show improvements. The initial experiments were carried out using a separate evaluation dataset.

---

[1] http://www-nlpir.nist.gov/projects/duc/index.html
[2] http://www.nist.gov/tac/

## 4.4 Baselines

Summaries generated from the MultiGBS algorithm were compared against LexRank and a new method based on BERT. Based on our previous research, the LexRank algorithm provided better results in comparison with other graph-based methods [41]. Therefore, we employed Lexrank as a representative of the graph-based methods for the initial experiments.

**LexRank** is a graph-based method that uses TF/IDF cosine similarity and generates a similarity matrix. Then, it extracts the most important sentences based on centrality. This idea of selecting meaningful sentences is similar to PageRank [32]. The second method, which we will call in the rest of this paper "**Leveraging BERT**", is based on a combination of BERT with K-Means clustering [61].

## 4.5 Experimental results

In this section, the results of different experiments are presented. Section 4.5.1 reports the results of initial experiments, and in Section 4.5.2 the proposed MultiGBS method is compared with other methods[3].

### 4.5.1 Initial experiments

The proposed method has two versions: MultiGBS Basic and MultiGBS Enhanced. MultiGBS Basic uses the information from the multi-layer graph to rank sentences and create a summary. In this step, the MultiGBS Basic algorithm used MetaMap to extract the concepts and MultiRank to sort the sentences (The results for comparing MultiRank and a simple baseline can be found in Appendix B). Therefore, for Q1, two different types of graphs were created: weighted and unweighted graphs. For unweighted graphs, the algorithm processed a pair of sentences that were linked to each other if they had a similarity above the threshold. We analyzed the different values of similarity measures to assign 0.1, 0.2, and 0.3 as the thresholds. For weighted graphs, the

---

[3] For all the tests, we employed the following runtime parameters:

"*metamap -L 2018 -Z 2018AA --lexicon db -V USAbase -E -ApGj+ --XMLf –E*" for MetaMap and "*semrep -Z 2018 -L 2006 -A -X –E*" for SemRep."

original weights without any threshold are considered. Table 1 shows the results of different ROUGE scores for these weighted and unweighted models.

Table 1: The ROUGE scores for weighted and unweighted models using MultiGBS Basic. (Compression rate=20%)

|  | LexRank | Unweighted | | | Weighted (No threshold) |
|---|---|---|---|---|---|
|  |  | Threshold=0.1 | Threshold=0.2 | Threshold=0.3 |  |
| **Recall** | | | | | |
| ROUGE-L | 0.279 | 0.287 | 0.282 | 0.281 | **0.294** |
| ROUGE-1 | 0.693 | 0.713 | 0.71 | 0.708 | **0.717** |
| ROUGE-2 | 0.348 | 0.351 | 0.351 | 0.353 | **0.353** |
| ROUGE-SU4 | 0.378 | 0.384 | 0.383 | **0.387** | 0.386 |
| **Precision** | | | | | |
| ROUGE-L | **0.112** | 0.091 | 0.093 | 0.098 | 0.094 |
| ROUGE-1 | **0.202** | 0.18 | 0.184 | 0.188 | 0.179 |
| ROUGE-2 | **0.107** | 0.093 | 0.096 | 0.099 | 0.092 |
| ROUGE-SU4 | **0.119** | 0.104 | 0.107 | 0.111 | 0.103 |
| **F-measure** | | | | | |
| ROUGE-L | **0.158** | 0.137 | 0.138 | 0.143 | 0.14 |
| ROUGE-1 | **0.306** | 0.282 | 0.287 | 0.291 | 0.281 |
| ROUGE-2 | **0.161** | 0.144 | 0.147 | 0.151 | 0.144 |
| ROUGE-SU4 | **0.178** | 0.160 | 0.164 | 0.169 | 0.160 |

As shown in Table 1, both the weighted and unweighted model of MultiGBS Basic achieve higher recall in comparison to LexRank, but for precision and F-measure, LexRank has better results. The best recall results are obtained by the weighted graph, which confirms the effect of employing domain-specific knowledge for extracting meaningful sentences. In this step, MultiGBS Enhanced is presented to improve the precision and F-measure. The recall results show that MultiGBS Basic extracts the expected concepts, but it also extracts irrelevant concepts, leading to low precision results. Based on Eqs. (4) and (5), we can control the number of meaningful concept to improve the precision of the results. Consequently, MultiGBS Enhanced computes the ratio of the number of concepts in each sentence to the total number of concepts in the text using Eq. (2) and ranks the sentences based on Eq. (3). (The comparison results of MultiGBS Enhanced and LexRank are presented in Table 2.)

To answer Q2, the MultiGBS Enhanced algorithm was evaluated with different weights for $\gamma$ and $\theta$ in Eq. (3) to show the effect of extracting irrelevant concepts. MultiGBS Enhanced needs to check the number of concepts; setting a high or low value for $\theta$ causes the loss of meaningful concepts (We evaluated high or low values of $\theta$ in Appendix C). Therefore, we selected $\theta$ in the range [-1, 1] to achieve the best F-measure results. The different ROUGE scores (R-1, R-2, R-W-1.2, and R-SU4) for all the summarizers are given in Table 2.

**Table 2**: The ROUGE scores of the summaries that MultiGBS Enhanced created with the different weights for γ and θ. (compression rate=20%)

| Evaluation Metrics | LexRank | MultiGBS Enhanced | | | | | | | |
|---|---|---|---|---|---|---|---|---|---|
| | | $\gamma$ = 1 | 1 | 1 | 1 | 1 | 1 | 1 | 1 |
| | | $\theta$ = 1 | −1 | 0.5 | −0.5 | 0.25 | −0.25 | 0.1 | −0.1 |
| | | **Recall** | | | | | | | |
| ROUGE-L | 0.279 | 0.291 | 0.248 | **0.294** | 0.261 | 0.293 | 0.269 | 0.293 | 0.282 |
| ROUGE-1 | 0.693 | 0.731 | 0.493 | **0.732** | 0.541 | **0.732** | 0.596 | 0.73 | 0.661 |
| ROUGE-2 | 0.348 | 0.361 | 0.19 | **0.363** | 0.221 | **0.363** | 0.259 | 0.363 | 0.305 |
| ROUGE-SU4 | 0.378 | 0.395 | 0.217 | **0.397** | 0.249 | **0.398** | 0.287 | 0.397 | 0.337 |
| | | **Precision** | | | | | | | |
| ROUGE-L | 0.112 | 0.08 | **0.148** | 0.082 | 0.14 | 0.083 | 0.128 | 0.086 | 0.111 |
| ROUGE-1 | 0.202 | 0.159 | **0.294** | 0.16 | 0.276 | 0.163 | 0.259 | 0.168 | 0.22 |
| ROUGE-2 | 0.107 | 0.082 | **0.123** | 0.083 | 0.121 | 0.085 | 0.12 | 0.088 | 0.107 |
| ROUGE-SU4 | 0.119 | 0.091 | **0.153** | 0.093 | 0.146 | 0.094 | 0.14 | 0.097 | 0.122 |
| | | **F-measure** | | | | | | | |
| ROUGE-L | 0.158 | 0.124 | **0.182** | 0.127 | 0.179 | 0.128 | 0.17 | 0.132 | 0.157 |
| ROUGE-1 | 0.306 | 0.256 | **0.358** | 0.259 | 0.356 | 0.262 | 0.352 | 0.269 | 0.323 |
| ROUGE-2 | **0.161** | 0.132 | 0.145 | 0.133 | 0.153 | 0.135 | 0.16 | 0.139 | 0.155 |
| ROUGE-SU4 | 0.178 | 0.146 | 0.174 | 0.148 | 0.179 | 0.15 | **0.183** | 0.154 | 0.175 |

The results confirm our hypothesis that irrelevant concepts can affect the results. When $\theta$ is negative, the precision values are better than other versions. Accordingly, MultiGBS Enhanced improves F-measure results by creating a multi-layer graph and reducing irrelevant concepts.

For Q3, different methods are used to extract the entities. First, the UMLS with MetaMap and OGER was employed. Second, BERN was used, which is based on BERT [62], which is a neural biomedical entity recognition and multi-type normalization tool [63]. Three different tools were evaluated based on 85 articles from the CRAFT corpus using the metrics precision, recall, and F1-measure [64]. OGER and MetaMap performed better than BERN, based on the results shown in Table 3.

**Table 3**: Comparison of the NER (Named Entity Recognition performance for MetaMap, OGER, and BERN)

|  | OGER | MetaMap | BERN |
| --- | --- | --- | --- |
| Precision | **0.4778** | 0.1635 | 0.3435 |
| Recall | 0.4964 | **0.4996** | 0.1514 |
| F-measure | **0.4819** | 0.2422 | 0.2026 |

Therefore, we incorporated OGER and MetaMap to create a multi-layer graph in the next evaluation. We created three different summaries and compared the ROUGE results for two different compression rates.

For Q4, we evaluated MultiGBS Enhanced with a different compression rate to show the effect of the size of the summary on the results. The typical size of the summary is between 15% and 35% of the size of the original text [41,43]. In general, adding more sentences as output increases the recall results. Consequently, it may be observed in Tables 1 and 2 that the best settings for the recall results were achieved by the MultiGBS Enhanced method. In this step, MultiGBS Enhanced used the outputs from OGER, MetaMap and a combination of MetaMap and OGER to extract entities and create three different summaries with higher recall ($\theta=0.5$ in Eq. (3)). The results are shown in Fig. 2, using compression rates=20% and 30%.

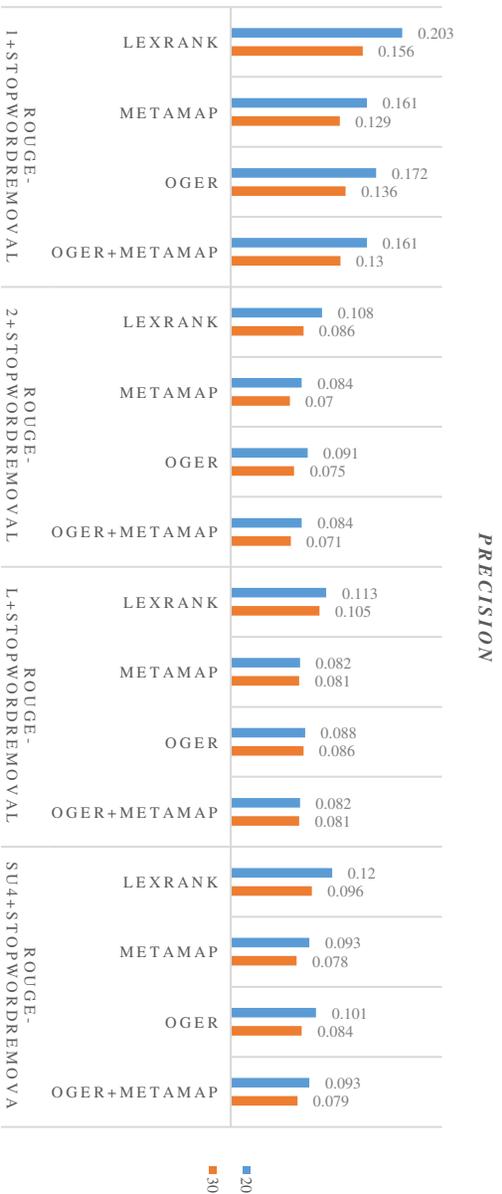

a) The recall results

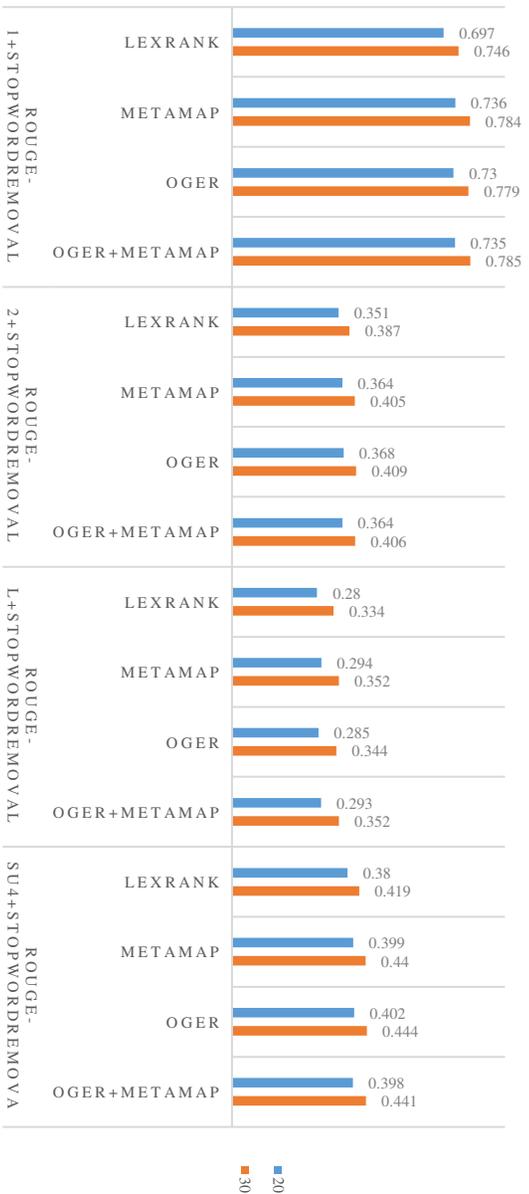

b) The precision results

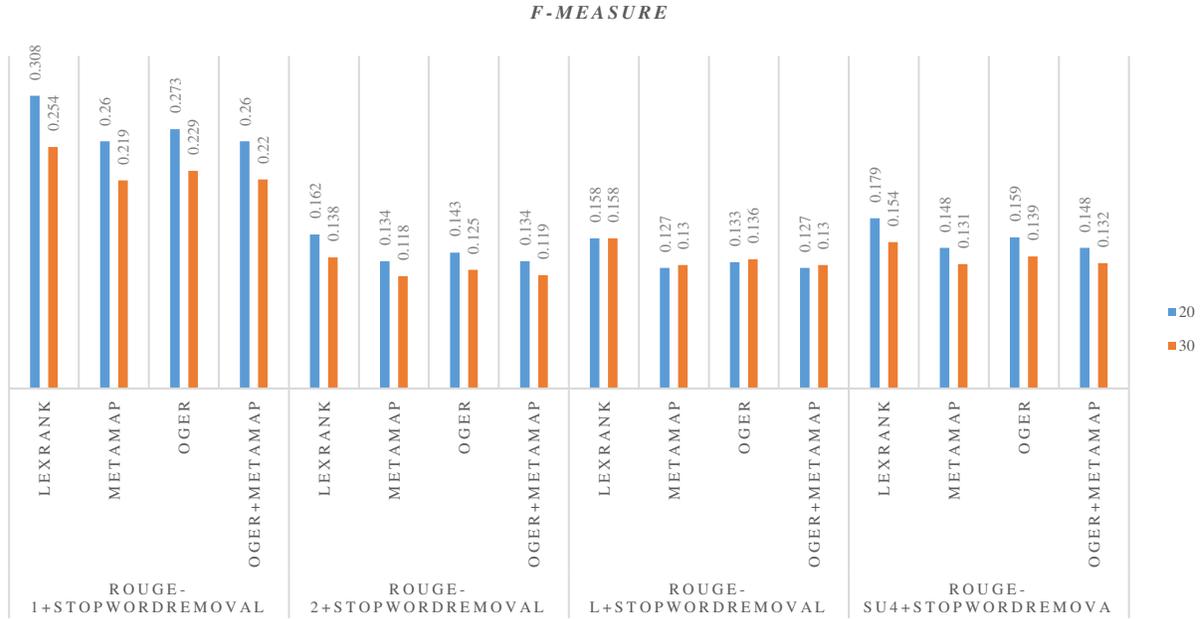

c)The F-measure results

**Fig. 2**: ROUGE scores for the summaries with compressions rate= 20% and 30%. MultiGBS Enhanced cerates three summaries with higher recall based on MetaMap, OGER and a combination of OGER and MetaMap.

As the results show, when the larger size of the summary is specified, the recall value increases (Fig. 2 (a)) but the precision and F-measure decrease (Fig. 2 (b) and (c)) for LexRank and all versions of MultiGBS. The results emphasize the importance of extracting meaningful concepts to improve the precision and F-measure results. Therefore, MultiGBS Enhanced need to decrease the effect of sentences with irrelevant concepts using the negative weight for θ.

It should be considered that documents of the same size could have abstracts of different sizes. For example, assume two papers with a size of 180 sentences, while the size of the abstract is 18 for the first paper and 32 for another. This situation appears in an identical way for our algorithm and for the other algorithms that we have considered in this study.

To answer Q5, we designed an ablation study and removed different layers for individual tests. We examined our algorithm with three layers and a subset of layers and reported the results in Table 4.

**Table 4**: The ROUGE results for comparing subsets of layers.

|  | 3 Layers(Semantic, Word, Co- reference) | | | 2 Layers(Semantic, Word) | | | 2 Layers(Semantic, Co-reference) | | | 2 Layers(Word, Co-reference) | | |
|---|---|---|---|---|---|---|---|---|---|---|---|---|
|  | Recall | Precision | F-measure | Recall | Precision | F-measure | Recall | Precision | F-measure | Recall | Precision | F-measure |
| ROUGE-L | **0.398** | **0.182** | **0.238** | 0.389 | 0.178 | 0.233 | 0.392 | 0.174 | 0.229 | 0.332 | 0.169 | 0.213 |
| ROUGE-1 | 0.614 | 0.232 | **0.322** | 0.606 | 0.230 | 0.318 | **0.617** | 0.216 | 0.305 | 0.543 | **0.239** | 0.315 |
| ROUGE-2 | **0.203** | **0.080** | **0.111** | 0.201 | 0.079 | 0.110 | 0.199 | 0.074 | 0.104 | 0.155 | 0.075 | 0.097 |
| ROUGE-SU4 | 0.268 | 0.111 | **0.150** | 0.267 | 0.109 | 0.148 | **0.272** | 0.103 | 0.143 | 0.228 | **0.111** | 0.141 |

As it is shown, using the graph with three layers as an enriched model of the documents produces better results. The number of abbreviations, acronyms, and symbols in the biomedical domain is higher than in other domains. The results confirm that using a multi-layer graph with different similarity measures that take the domain characteristics into account is a high-quality approach for extracting semantic concepts and meaningful sentences.

### 4.5.2 Summary comparison

To evaluate the performance of the proposed summarization system, we compared the results of "Leveraging BERT", LexRank, and our algorithm based on ROUGE and BERTScore with a compression rate = 20%. The results are shown in Table 5. We employed a Wilcoxon signed-rank test to show the statistical significance of the results. In this step, if the p-value is lower than or equal to 0.05, the differences between two paired samples are statistically significant. All cases are significantly different except the F-measure of ROUGE-2 for Leveraging BERT compared with MultiGBS which is shown with a + in Table 5. (Appendix D presents the P-value comparison of MultiGBS and other baselines.)

The F1-measure is high if both precision and recall are high. Consequently, MultiGBS takes advantage of a multi-layer graph to cover the different relationships in combination with the number of concepts to improve the F1-measure.

**Table 5**: The ROUGE and BERTScore results for the compared methods. A "+" appears in cases where the MultiGBS does not perform significantly differently.

|  | Leveraging BERT | LexRank | MultiGBS Enhanced |
|---|---|---|---|
| | **Recall** | | |
| ROUGE-L | 0.403 | **0.559** | 0.452 |
| ROUGE-1 | 0.649 | **0.720** | 0.586 |
| ROUGE-2 | 0.176 | **0.256** | 0.151 |
| ROUGE-SU4 | 0.345 | **0.425** | 0.301 |
| Bert Score | 0.834 | **0.836** | 0.833 |
| **Precision** | | | |
| ROUGE-L | 0.062 | 0.083 | **0.103** |
| ROUGE-1 | 0.101 | 0.064 | **0.111** |
| ROUGE-2 | 0.034 | 0.027 | **0.035** |
| ROUGE-SU4 | 0.045 | 0.033 | **0.052** |
| BertScore | 0.777 | 0.778 | **0.784** |
| **F-measure** | | | |
| ROUGE-L | 0.095 | 0.129 | **0.146** |
| ROUGE-1 | 0.155 | 0.108 | **0.164** |
| ROUGE-2 | 0.051 | 0.0455 | **0.052+** |
| ROUGE-SU4 | 0.070 | 0.057 | **0.075** |
| BertScore | 0.804 | 0.805 | **0.806** |

The results show that MultiGBS improves the F-measure. The precision value is more likely to measure the redundancy in the summary, and the recall estimates the coverage. F-measure is a metric that reflects both the coverage and redundancy. Therefore, a summarizer with a higher F-measure value would be able to extract meaningful sentences with less repetition. The results confirm that the integration of domain-specific knowledge and multi-layer graph modelling creates better summaries.

## 5. Conclusion

In this paper, we have presented MultiGBS as a multi-layer graph model for biomedical text summarization. Every layer in MultiGBS represents a different aspect or similarity measure for the elements of the input document to cover the different relationships among sentences. This enriched model provides more room for improved summarization. The method was evaluated based on a collection of papers and compared with different methods such as LexRank and Leveraging BERT. The results show that the proposed summarizer outperforms the other tools according to the metrics ROUGE and BERTScore.

Further research in this area could focus on more similarity measures depending on the context of the documents. Moreover, the user requirements may be defined as a complement to the multi-layer graph model.

# Appendix A: Compare three different summaries

Below we show summaries generated by different methods for the paper "Voltage-*dependent anion channel (VDAC) is involved in apoptosis of cell lines carrying the mitochondrial DNA mutation*"( PMCID: PMC2779793, PMID: 19895710)

**Abstract (reference summary):**

*"Background: The mitochondrial voltage-dependent anion channel (VDAC) is increasingly implicated in the control of apoptosis. We have studied the effects the mitochondrial DNA (mtDNA) tRNAIle mutation on VDAC expression, localization, and apoptosis.*

*Methods: Lymphoblastoid cell lines were derived from 3 symptomatic and 1 asymptomatic members of a family with hypertension associated with the A4263G tRNAIle mutation as well as from control subjects. Mitochondrial potential ($\Delta\Psi m$) and apoptosis were measured by flow cytometry; co-localization of VDAC and Bax was evaluated by confocal microscopy.*

*Results: Expression of VDAC and Bax in mtDNA cell lines was found to be increased compared to controls, while expression of the small conductance calcium-dependant potassium channel (sKCa) was unchanged. Confocal imaging revealed co-localization of VDAC/Bax on the outer mitochondrial membrane of A4263G cell lines but not from controls. Flow cytometry indicated that the mitochondrial potential was decreased by 32% in mutated cells versus controls while rates of apoptosis were increased (P < 0.05). The difference was attenuated by Cyclosporin A (CsA, 2 μM), a blocker of VDAC.*

*Conclusion: We conclude that increased expression of mitochondrial VDAC and subcellular co-localization of VDAC/Bax increases mitochondrial permeability and apoptosis in cell lines carrying the mtDNA tRNAIle A4263G mutation*."

**LexRank Output:**

*"Multiple environmental and genetic factors are known to predispose to essential hypertension, with genetic predisposition contributing to 30-60% of the pathoetiology of the disease [1]. To study the link between the mitochondrial tRNAIle A4263G mutation, high blood pressure, and apoptosis we established lymphoblastoid cell lines from individuals carrying the A4263G mutation and from controls. We report that the A4263G mutation is associated with changes in VDAC expression, localization, and levels of apoptosis.Cell lines derived from 4 members of the Chinese family with a maternally-inherited predisposition to hypertension.To evaluate the effects of CsA on delta psim, cell lines were pre-incubated with CsA (Novartis, 2 M) for 30 min and washed 5\* in PBS before measurements. As shown in Figure 2, the A4263G mutation was associated with a significant (P < 0.05) increase in the levels of expression of both VDAC and Bax. After incubation with CsA for 16 h, VDAC mRNA levels decreased significantly and even lower than the control levels (P < 0.05) while there was no significant change in the levels of either Bax or sKCa mRNA (Figure 2).VDAC was localized on the outer membrane of the mitochondrial, but Bax was expressed in*

*the cytoplasm, transferred to outer membrane of mitochondrial and combined to VDAC under pathological state. In contrast, in cell lines from 2 tRNAIle A4263G subjects (III-14 and III-19) co-localization of VDAC and Bax was revealed by intense yellow fluorescence.We also investigated the effects of CsA on co-localization. Confocal imaging revealed that CsA treatment abolished co-localization of VDAC and Bax.The magnitude of delta psim is controlled by the activity of the mitochondrial KATP (mitoKATP) and mitoKCa channels, as well as by the mitochondrial permeability transition pore (PTP) [21].After incubation with CsA for 30 min the delta psim of both control and the mutated cell lines was increased; the increase was 33.6% in control cells but was 84.4% in III-14 mutant cells and 137.7% in III-19 mutant cells (Figure 4).We then compared delta psim values between cell lines from controls and subjects carrying the A4263G mutation, both prior to and following incubation with CsA. CsA pretreatment increased delta psim in all cell lines, and after CsA treatment there was no significant difference between delta psim values in control and mutant cells (Figure 5).To determine whether changes in VDAC mRNA levels, localization, and delta psim values in lines from subjects carrying the A4263G mutation are associated with changes in apoptosis, cells were analyzed using Annexin V-FITC and flow cytometry, a sensitive assay for apoptosis. mtDNA sequence analysis revealed a A4263G mutation in the mtDNA tRNAIle gene that is extraordinarily conserved from bacteria to human.VDAC also appears to be an anchor point for pro- and anti-apoptotic proteins; it has been hypothesized that VDAC contributes to increase of mitochondrial permeability that is involved in the initiation of apoptosis.We studied the effects of the mtDNA tRNAIle A4263G mutation on VDAC function and apoptosis. These changes were accompanied by a decrease in the mitochondrial membrane potential delta psim and increased levels of apoptosis versus control cell lines. The role of VDAC in these processes was confirmed by application of the selective VDAC inhibitor Cyclosporin A (CsA): the inhibitor abolished co-localization with Bax, restored delta psim to levels of control CsA-treated cells, and decreased levels of apoptosis. Taken together, these results suggest that the mtDNA A4263G mutation, inferred to cause mis-charging of tRNAIle and consequent amino acid substitutions in mitochondrial proteins, may exert its pro-hypertensive effects by deregulating the expression of VDAC, in turn leading to increases in programmed cell death.So we conclude that increased expression of mitochondrial VDAC and subcellular co-localization of VDAC/Bax increases mitochondrial permeability and apoptosis in cell lines carrying the mtDNA tRNAIle A4263G mutation."*

**Leveraging BERT Output:**

*"Hypertension is an established risk factor for coronary heart disease, stroke, congestive heart failure and renal dysfunction, and is the major modifiable risk factor of poor prognosis in a variety of cardiovascular diseases. Importantly, the change was inferred to influence amino acid charging of tRNA [6] and was therefore likely to lead to translational amino-acid substitutions in some mitochondrial proteins, with major consequences for mitochondrial function. Recent studies have reported that the outer mitochondrial membrane voltage-dependent anion channel (VDAC) is associated with type 2 diabetes mellitus [7, 8, 9], an important finding in view of the link between diabetes and mitochondrial function [10, 11]. We report that the A4263G mutation is associated with changes in VDAC expression, localization, and levels of apoptosis. RNA (2 g) was treated with ribonuclease-free deoxyribonuclease and cDNA was synthesized using Moloney murine leukemia virus reverse transcriptase (Invitrogen Life Technologies,*

*Carlsbad, CA); cDNA (2 l) was subjected to 44 cycles of PCR amplification, generating a single specific amplification product of the expected size. Cells were grown on glass coverslips in 6-well plates. Comparison of continuous variables was performed using the unpaired Student's t test. Statistical significance was set at $P < 0.05$. After incubation with CsA for 16 h, VDAC mRNA levels decreased significantly and even lower than the control levels ($P < 0.05$) while there was no significant change in the levels of either Bax or sKCa mRNA (Figure 2). VDAC was localized on the outer membrane of the mitochondrial, but Bax was expressed in the cytoplasm, transferred to outer membrane of mitochondrial and combined to VDAC under pathological state. Localization was imaged under confocal microscopy. Representative sections are shown in Figure 3. Cell lines were incubated with CsA for 30 min prior to analysis. CsA pretreatment increased delta psim in all cell lines, and after CsA treatment there was no significant difference between delta psim values in control and mutant cells (Figure 5). To determine whether changes in VDAC mRNA levels, localization, and delta psim values in lines from subjects carrying the A4263G mutation are associated with changes in apoptosis, cells were analyzed using Annexin V-FITC and flow cytometry, a sensitive assay for apoptosis. Levels of apoptosis were increased by 30% in cell lines carrying the mutation (Figure 6, left panel). The voltage-dependent anion channel (VDAC) is a highly conserved protein located on the outer mitochondrial membrane [11]. VDAC also appears to be an anchor point for pro- and anti-apoptotic proteins; it has been hypothesized that VDAC contributes to increase of mitochondrial permeability that is involved in the initiation of apoptosis. We studied the effects of the mtDNA tRNAIle A4263G mutation on VDAC function and apoptosis. Although the exact mechanism is unknown, it may be associated with dysfunction of energy metabolism and consequent increase in the levels of reactive oxygen species (ROS) [31]."*

**MultiGBS Output:**

*"Multiple environmental and genetic factors are known to predispose to essential hypertension, with genetic predisposition contributing to 30-60% of the pathoetiology of the disease [1].We report that the A4263G mutation is associated with changes in VDAC expression, localization, and levels of apoptosis. Dual immunostaining was used to asssess the co-localization of Bax and VDAC-1 polypeptides. As shown in Figure 2, the A4263G mutation was associated with a significant ($P < 0.05$) increase in the levels of expression of both VDAC and Bax. Localization was imaged under confocal microscopy. Separate VDAC (green) and Bax (red) fluorescence was detected in control cell lines, indicating that the respective polypeptides were separately localized. As shown in Figure 5, delta psim values were significantly decreased ($P < 0.05$) in both cell lines harboring the mutation. Levels of apoptosis were increased by 30% in cell lines carrying the mutation (Figure 6, left panel). The voltage-dependent anion channel (VDAC) is a highly conserved protein located on the outer mitochondrial membrane [11]. VDAC, in association with ANT (adenine nucleotide translocator), mediates the transport of ATP and ADP both into and out of the mitochondrion [11]. VDAC also mediates mitochondrial $Ca2+$ and may play a key role in intracellular $Ca2+$ signaling [28]. We report that expression level of VDAC mRNA in cell lines carrying the mutation was significantly decreased. We hypothesis that high apoptosis of the mutated cell lines was associated with increased expression of VDAC mRNA, which could be inhibited by the CsA, a kind of inhibitor of VDAC. In addition, imaging revealed that VDAC was co-localized with Bax protein in cell lines carrying the 4263 AG mutation.These changes were accompanied by a decrease in the*

*mitochondrial membrane potential delta psim and increased levels of apoptosis versus control cell lines. Finally this article we conclude that increased expression of mitochondrial VDAC and subcellular co-localization of VDAC/Bax increases mitochondrial permeability and apoptosis.*"

## Appendix B: Compare Multi Rank and simple baseline

We needed to aggregate different similarity measures without losing data. The simple weighted average algorithms don't consider the influence of each layer and lost data. The MultiRank algorithm discovers the centrality of nodes and layers based on different properties. To compare Multi Rank with a simple baseline, we created three different graphs based on similarity measures and employed the page rank algorithm to rank each graph separately. Then a simple weighted average is used to rank sentences to create summary. The results are shown in Table B.1

**Table B.1**: The recall results for comparing Multi Rank with simple baseline

|            | Simple Baseline | MultiRank |
|------------|-----------------|-----------|
|            | Recall          |           |
| ROUGE-L    | 0.440           | **0.457** |
| ROUGE-1    | 0.696           | **0.711** |
| ROUGE-2    | 0.290           | **0.308** |
| ROUGE-SU4  | 0.359           | **0.375** |

The results indicate, the Multirank has a better performance than the simple baseline.

# Appendix C: The high and low values of θ

We selected a subset of train data and assigned different values (very low and very high) .The results are shown in Table C.1

Table C.1: ROUGE scores for different values of θ.

|           | θ=-1  | θ=-10 | θ=10  |
|-----------|-------|-------|-------|
|           | **Recall** | | |
| ROUGE-L   | 0.41  | 0.333 | 0.519 |
| ROUGE-1   | 0.644 | 0.482 | 0.8   |
| ROUGE-2   | 0.268 | 0.153 | 0.412 |
| ROUGE-SU4 | 0.322 | 0.204 | 0.462 |
|           | **Precision** | | |
| ROUGE-L   | 0.244 | 0.25  | 0.158 |
| ROUGE-1   | 0.318 | 0.38  | 0.172 |
| ROUGE-2   | 0.136 | 0.129 | 0.091 |
| ROUGE-SU4 | 0.171 | 0.185 | 0.103 |
|           | **F-measure** | | |
| ROUGE-L   | 0.301 | 0.28  | 0.24  |
| ROUGE-1   | 0.414 | 0.412 | 0.279 |
| ROUGE-2   | 0.176 | 0.136 | 0.146 |
| ROUGE-SU4 | 0.217 | 0.188 | 0.166 |

# Appendix D: Statistical test and histogram of distribution of data

We employed a Wilcoxon signed-rank test to show the statistical significance of the results based on distribution of data. The distribution of data for all ROUGE scores is shown in Fig C.1

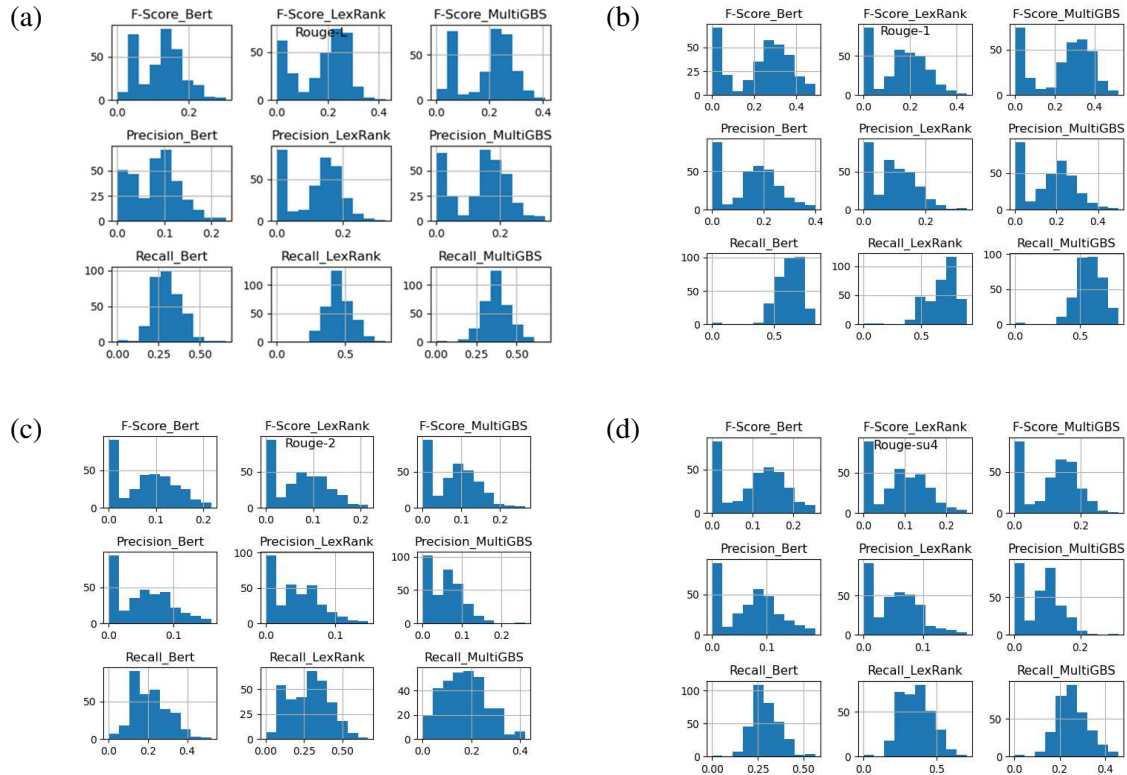

**Fig C.1** Distributions of data for (a): ROUGE-L, (b): ROUGE-1, (c): ROUGE-2 and (d): ROUGE-SU4

Table C.1 shows P-value comparison of MultiGBS and other baselines.

**Table C.1**: P-value comparison of MultiGBS and other baselines. A "*" appears in cases where the MultiGBS does not perform significantly differently.

|  | **Leveraging BERT & MultiGBS** | | | **LexRank & MultiGBS** | | |
|---|---|---|---|---|---|---|
|  | **Recall** | **Precision** | **F-measure** | **Recall** | **Precision** | **F-measure** |
| ROUGE-L | 6.26E-32 | 2.08E-51 | 6.40E-51 | 7.61E-39 | 2.18E-31 | 3.20E-10 |
| ROUGE-1 | 1.23E-32 | 3.88E-14 | 8.11E-08 | 3.54E-49 | 5.34E-54 | 2.49E-53 |
| ROUGE-2 | 7.63E-20 | 0.023203 | * | 2.64E-47 | 1.44E-28 | 5.09E-09 |
| ROUGE-SU4 | 8.01E-30 | 1.04E-23 | 1.82E-10 | 6.42E-52 | 9.72E-52 | 2.40E-42 |